\documentclass[aps,groupedaddress,twocolumn]{revtex4}

\usepackage{epsf}
\usepackage{epsfig}

\begin{document}
%\draft

\title {
\small published in J. Electron Spectroscopy and Related Phenomena \textbf{183},48--52(2011) \\[1.0cm]
%\large
Carbon K-shell Photo Ionization of CO: \\
Molecular frame angular Distributions of normal and conjugate
shakeup Satellites}

\author{T. Jahnke$^1$}
\author{J. Titze$^1$}
\author{L. Foucar$^1$}
\author{R. Wallauer$^1$}
\author{T. Osipov$^2$}
\author{E. P. Benis$^3$}
\author{O. Jagutzki$^1$}
\author{W. Arnold$^1$}
\author{A. Czasch$^1$}
\author{A. Staudte$^4$}
\author{M. Sch\"offler$^2$}
\author{A. Alnaser$^2$}
\author{T. Weber$^2$}
\author{M. H. Prior$^2$}
\author{H. Schmidt-B\"ocking$^1$}
\author{R. D\"orner$^1$}
\email{doerner@atom.uni-frankfurt.de}

\affiliation{
$^1$ Institut f\"ur Kernphysik, University of Frankfurt,  Max-von-Laue-Str. 1,  D-60438 Frankfurt, Germany\\
$^2$ Lawrence Berkeley National Laboratory,   Berkeley CA 94720\\
$^3$ Department of Physics, University of Ioannina, GR-45110 Ioannina, Greece\\
$^4$ Steacie Institute for Molecular Sciences, NRC, 100 Sussex Drive, Ottawa, Ontario, Canada}

\date{\today}

\begin{abstract}
We have measured the molecular frame angular distributions of
photoelectrons emitted from the Carbon K shell of fixed-in-space
CO molecules for the case of simultaneous excitation of the
remaining molecular ion. Normal and conjugate shake up states are
observed. Photo electrons belonging to normal $\Sigma$-satellite
lines show an angular distribution resembling that observed for
the main photoline at the same electron energy. Surprisingly a
similar shape is found for conjugate shake up states with
$\Pi$-symmetry. In our data we identify shake rather than electron
scattering (PEVE) as the mechanism producing the conjugate lines.
The angular distributions clearly show the presence of a $\Sigma$
shape resonance for all of the satellite lines.
\end{abstract}
% insert suggested P in braces on next line
%\pacs{33.80.Eh., 33.90.+h}
\maketitle

% body of paper here
\section{Introduction}
The present paper combines two vibrantly discussed topics:
molecular frame angular distributions of innershell photoelectrons
and few body processes induced by a single photon. We will ask how
molecular frame angular distributions are influenced by the
involvement of other active electrons and in turn what we can
learn about few electron processes from molecular frame angular
distributions. Before we discuss our experiment we give a brief
summary on the background of both topics.

\subsection{Photon induced many electron transitions}
Photoionization of atoms and molecules leads in many cases to a
simultaneous excitation of the remaining ion
\cite{Carlson65pra,Aberg67}. Such ionization excitations manifest
themselves in satellites lines to the main photoelectron line in
the electron energy distribution. As this process is inherently a
many electron process its description becomes particulary
challenging: a widely used two-step model for ionization
excitation implies that either the photon leads to the emission of
the photoelectron by a primary \emph{dipole} transition followed
by a \emph{monopole} transition inside the molecular ion (normal
shake up). Alternatively the photon may excite the molecule in a
\emph{dipole} transition and subsequently an electron is shaken
off to the continuum in a \emph{monopole} transition (conjugate
shake up). At high excess energies conjugate shake up dies off and
only molecular ion states with the same symmetry as the ground
state are populated. Conjugate shake up however contributes
significantly at low electron energies.

The word "shake up" was introduced in the context of the sudden
approximation. It refers to a {\em "mechanism"} by which the
excitation proceeds. It can be formalized as a Feynman-diagram in
many body perturbation theory \cite{Hino93,Mcguire95,Dalgarno92}.
In an intuitive picture of that description the photon induces a
sudden single active electron transition in which it gives its
angular momentum either to the ejected or the excited electron.
This ionization or excitation step changes the potential and hence
the eigenstates of the system. As the wave functions of the
remaining electrons relax to the new eigenstates of the altered
potential, they have a non zero overlap to excited or even
continuum states to which they are "shaken up" or "shaken off".
This simple picture, however, does not answer the question how the
photon energy is partially transferred from the first electron,
which absorbed the photon, to the shake electron, but it permits
calculation of the shake probabilities as simple overlap integrals
without an operator \cite{Aberg67}.

Within the framework of many body perturbation theory a second
mechanism of electron-electron scattering, which actually has a
classical analog, can lead to a two electron transition upon
absorption of a single photon. This mechanism comes at a variety
of names like TS1 \cite{Mcguire95}, knock-off
\cite{Schneider02prl,Schneider03pra} or PEVE (photoelectron
valence electron interaction)
\cite{Liu08prl,Pavylchev99jpb,Defanis02prlb}. It is also related
to the rescattering mechanism in the multiphoton context
\cite{Becker05jpb}. The wording mechanism is certainly a
simplification as the depicted scenario describes contributions
to a transition amplitude. The details of these amplitudes have
been worked out in great detail for the two electron process of
single photon double ionization of helium \cite{Knapp02prl}.

Similar to the shake up mechanism, PEVE can be "normal" or
"conjugate" which means that it can either lead to the same symmetry
or to a different symmetry of the molecular ion compared to that of the
main line. Such a change of symmetry is possible since the
electrons can exchange angular momentum in the internal collision.
For the case of a conjugate process PEVE and shake can be
experimentally separated by examining the orientation of the
molecular axis to the polarization \cite{Liu08prl}: if e.g. in the
conjugate shake up the photon induces an excitation from
a $\Sigma$ to a $\Pi$ orbital the molecular axis has to be
perpendicular to the polarization during the photo absorption.

\subsection{Molecular frame angular distributions}
Angular distributions of photoelectrons emitted from molecules
exhibit a very rich structure in the laboratory frame - if the
molecular axis is fixed in space\cite{Kaplan69,Dehmer75prl}. What is the physical origin of this
structure? From the perspective of angular momentum one may say
that the outgoing photoelectron wave is a coherent superposition
of several angular momentum states, even within the dipole
approximation where only 1$\hbar$ is deposited by the photon into
the system. These high angular momenta in the continuum electron
wave function are compensated by a rotational excitation of the
molecular ion left behind, which can also be measured directly
\cite{Choi94prl}. One may say that the photoionization process splits the
many body wave function of electrons and nuclei of the neutral
molecule into an angular momentum entangled wave function of a
free electron carrying angular momentum in the angular
distribution and a molecular ion carrying the equivalent angular
momentum in rotation. A simplistic and mechanistical way to
understand the creation of these high angular momentum states is
suggested by the multiple scattering picture: a dipolar
photoelectron wave is created in the K-shell of the molecule. On
the way through the molecule this wave is multiply scattered at
the multi center potential of the molecule. It is this multiple
scattering which creates the angular momentum in the rotational
degrees of freedom of the molecular ion and in the electron wave
itself. In a particle picture the minima and maxima in the
electron angular distribution are then interpreted as destructive
or constructive interference between different pathways of the
electron through the molecular environment. In a wave formulation
they are the interference pattern of the direct dipolar
electron wave with all multiply scattered electron waves.
Such multiple scattering effects can be employed to use the photoelecton angular distribution to illuminate a molecule from within
\cite{Landers01prl}.
Particularly rich patterns arise when the electron wave resonates
inside the structure of the multicentric potential giving rise to a
\emph{shape resonance} \cite{Dehmer75prl,Shigemasa95prl}. For homonuclear molecules, there is a
second mechanism which gives rise to higher order angular momenta: the coherent emission of the primary electron wave from two or
more centers. Even if multiple scattering is small, as it is e.g.
the case for protons as scatterers in H$_2$, this potentially
delocalized nature of the photoeffect \cite{schoeffler08science}  alone can
create diffraction structures in the angular distributions \cite{Kaplan69,Akoury07science,Fernandez07prl,Kreidi08prl,schoeffler08pra}
and hence also rotationally excite the ions.

\subsection{Carbon K-shell ionization of CO}
In the remainder of this paper we have chosen the example of
Carbon K-shell ionization of fixed in space CO to investigate the
interplay between creation of satellite lines via excitation with
the molecular frame angular distributions. It is certainly
tempting to try to understand the angular distribution of the
satellites as being created by two steps: first an initial
$p$-wave or $s$-wave of an outgoing electron is created by  shake
or conjugate shake process respectively and in a second step this
low angular momentum wave is multiply scattered in the molecular
potential where higher angular momenta are added. This simplistic
picture, however, neglects the molecular symmetry of the wave
functions. Our data show that an alternative view, also put
forward in \cite{Ueda05prl} is more
appropriate. The $s$-wave character for the shaken off electron in
the atomic case refers to the fact that no angular momentum is
transferred from the photon to the shaken off electron. For the
molecular case, where $s$ is not a good quantum number, that idea
can be generalized to shake up in which the the $\Sigma$ and $\Pi$
character of the bound state wave function is conserved upon the
transition to the continuum.

The Carbon K satellite spectrum of CO has been studied in detail
experimentally
\cite{Hemmers95jpb,ungier84prl,randall93prl,reimer86prl,reich94pra,
Ueda05prl,Ehara06jcp} and theoretically
\cite{Angonoa87jcp,Bandarage93pra,Schirmer91pra,Schirmer87jpb,Ehara06jcp}. Two major satellite features S0
and S1 corresponding to singlet and triplet coupled excitations
have been observed \cite{reimer86prl} and calculated
\cite{Schirmer91pra}. They show a very different photon energy
dependence. High resolution studies have then resolved the various
states contributing to these two main features (see Fig. 1 below
and Table 1). Hemmers \emph{et al.} \cite{Hemmers95jpb} have
measured the $\beta$-parameters of the electron for these lines.
They found that - as expected by the simple picture above - the
satellite lines fall into two groups: those with $\beta=0$ which
are conjugate shake up satellites and those whose $\beta$ is
similar to the one of the main line which are direct (i.e. normal
shake up) satellites. In our study presented here we measure the
molecular ion angular distribution in addition. As outlined above
this will clearly allow us to also pin down not only the symmetry
(normal or conjugate) but also the excitation mechanism (shake
versus PEVE). As a further step we will then investigate the
electron angular distributions in the body fixed frame. This will
in particular allow identification of possible shape resonances.

\section{Experiment}
The experiment has been performed at beamline BL4 of the Advanced
Light Source at Lawrence Berkeley National Laboratory using the
COLTRIMS technique \cite{doerner00pr,ullrich03rpp,Jahnke04JESP}.
The photon energies employed where in a region of 290~eV to
320~eV, in all cases the light was linearly polarized. We have set
the monochromator to a resolution of about 100~meV, sufficient to
resolve the major features of the satellite spectrum. The photon
beam is crossed at right angle with a supersonic CO beam. The
photoelectron and both ions are guided by weak electric
(12.3~V/cm) and magnetic (8~G) fields towards two large area
position and time sensitive micro channel plate detectors equipped
with delayline position readout \cite{jagutzki02nim,roentdek}.

In most cases the K shell ionization is followed by the emission of an
Auger electron. The doubly charged molecular ion dissociates. For
the further analysis we have selected only events where both ions
were detected and the kinetic energy release was above
10.2~eV. In this case the dissociation proceeds along steeply
repulsive curves and is rapid enough so that the molecular ion does
not rotate between the photoelectron ejection and the
fragmentation \cite{Weber01jpb,Weber03prl}: the axial recoil approximation is valid. In that case the orientation
of the molecule in the laboratory frame can be obtained. The two singly charged ions
are emitted back-to-back in a Coulomb explosion. Therefore, by measuring the momenta of these two fragments
the direction of the molecular axis at the instant of photo ionization is determined. During the offline
analysis the measured electron momenta can be transformed from the laboratory frame
to the molecular frame. Furthermore, a subset of the data where the molecular axis
was oriented, for example perpendicularly, can be selected and plotted. That way molecular frame
electron angular distributions from fixed in space molecules are obtained from an
initially unoriented ensemble of molecules.

\section{Results}
In Figure 1(a) we show the measured electron energy distribution
at a photon energy of 318.97~eV in the region where satellite
lines occur. The line positions with the assignment are taken from
\cite{Hemmers95jpb} and are given in Table 1. In
\cite{Hemmers95jpb} the $\beta$ parameters of the electrons for
the respective satellite states are measured. On this basis
conjugate shake up states are identified by their isotropic
angular distribution of the electron in the laboratory frame.
Hemmers \emph{et al.} took this as an indication that the electron
results from a monopole transition. We list the known
configurations for the lines in Table 1 together with the
assignment of their symmetry and their suggested type of shake up
(normal or conjugate).

\begin{table}
\begin{tabular}{|c|c|c|c|}
  \hline
  Peak & Assignment & Binding     & type \\
       &            & energy (eV) &      \\
  \hline
  0 & $2\sigma^{-1}\  ^2\Sigma^+                                   $& 296.20 & Main \\
  1 & $2\sigma^{-1} 5\sigma^{-1} 2\pi^1(S^{'}=1) ^2\Pi             $& 304.10 & conjugate \\
  2 & $2\sigma^{-1} 1\pi^{-1} 2\pi^1(S^{'}=1) ^2\Sigma^+          $ & 304.85 & normal \\
  3 & $2\sigma^{-1} 1\pi^{-1} 2\pi^{1}~^{2}\Delta, {^2}\Sigma^-         $& 306.31 & conjugate \\
  4 & $2\sigma^{-1} 4\sigma^{-1}2\pi^{1}~{^2}\Pi                           $& 308.97 & conjugate \\
  5 & $2\sigma^{-1} 1\pi^{-1} 2\pi^1(S{'}=0) ^2\Sigma^-          $ & 311.29 & normal \\
  6 & unknown                                                   & 313.30 & conjugate \\
  7 & $2\sigma^{-1} 5\sigma^{-1} 6\sigma^{1} (S{'}=1) ^2\Sigma^+ $  & 313.97 & normal \\
  8 & $2\sigma^{-1}  5\sigma^{-1} 6\sigma^{1} (S{'}=0) ^2\Sigma^+$  & 315.37 & normal \\
  9 & $2\sigma^{-1}  5\sigma^{-1} 7\sigma^{1} (S{'}=1) ^2\Sigma^+$  & 316.25 & normal \\
  10 & unknown                                                  & 317.14 & normal \\
  \hline
\end{tabular}
\caption{CO C 1s ionized ground and excited electronic states
(from \cite{Hemmers95jpb}). The assignment of the type of shake up
state is taken from the $\beta$ parameter of the photoelectron
from \cite{Hemmers95jpb}. States assigned as "conjugate" are those
for which a $\beta$ close to zero has been observed.}
\end{table}

\begin{figure}[htbp]
 \begin{center}
  \epsfig{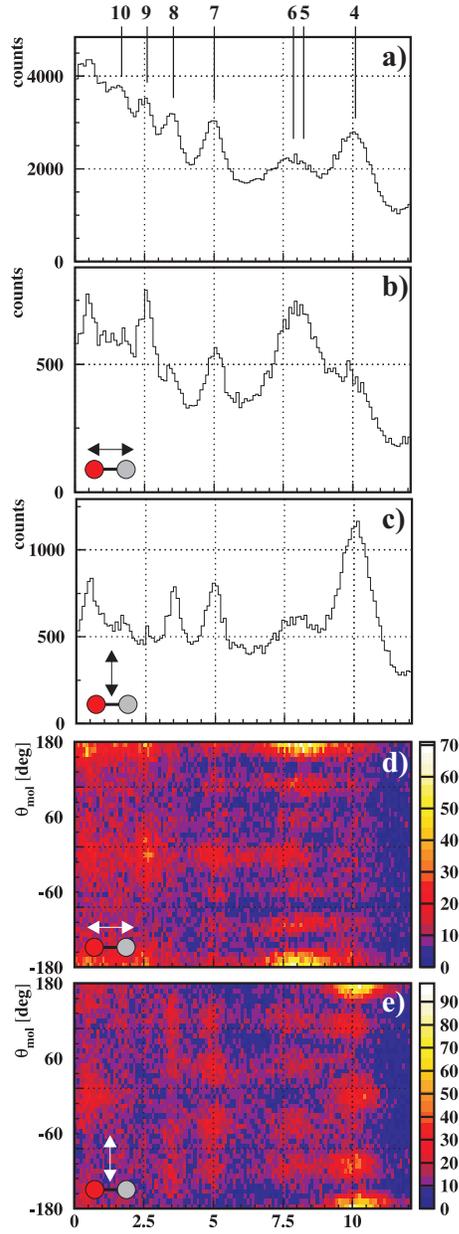}
 \end{center}
\caption{C K-shell ionization of CO at a photon energy of 318.97~eV.
(a) electron energy distribution integrated over all electron and
all ion emission angles. (b) as (a) but for molecules being oriented
parallel to the polarization. (c) same but for molecules being aligned
perpendicularly to the polarization. (d) electron energy versus angle
of the electron in the body fixed frame of the molecule, zero
degree is the direction of the Oxygen. Molecule and
polarization vector of the light are in parallel. (e) same as (d) but for molecules that are oriented
perpendicularly to the polarization.}
 \label{Fig1}
\end{figure}

The ion electron coincidence allows us to split our data set into
two subsets. The first one consists of events where the
orientation of the molecule is parallel, the second one includes
only events where the orientation is  perpendicular to the
polarization. These are shown in Fig. 1 (b),(c). Clearly, line
number 6 is almost missing for parallel orientation. This shows
that the mechanism responsible for the excitation of the conjugate
configuration is excitation by photon absorption followed by shake
up and not PEVE. This is different to the case of CO$_2$, where
satellites have been reported in \cite{Liu08prl} which - despite
of being of conjugate type - are populated mainly when the
molecule and the polarization vector of the light are in parallel.

\begin{figure}[htbp]
 \begin{center}
  \epsfig{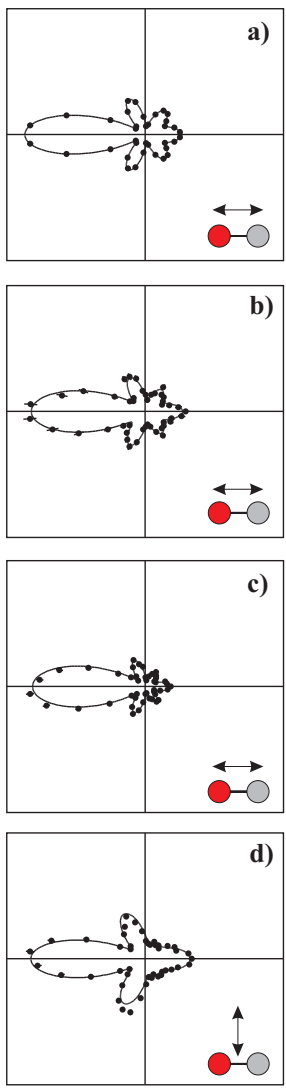}
 \end{center}
\caption{Angular distribution of the main line (a) the normal
satellite lines 5 (b) and 7 (c) and the conjugate line 4 (d) (see
Table 1 for numbering of the lines). For (a), (b) and (c) the
polarization is parallel to the molecular axis, for (d) it is
perpendicular. The lines show fits of Legendre polynomials to
guide the eye. The photon energy is 306.0~eV, 323.97~eV,
321.29~eV, 318.97~eV in panels (a), (b) , (c) , (d) chosen such
that the electron energy is about 10~eV in all four cases. The
icons show the orientation of the molecule and the polarization
vector of the light; the Oxygen atom is on the left.}
 \label{Fig2}
\end{figure}

The two color panels (d),(e) in Fig. 1 show the angular
distribution in the molecular frame for parallel and perpendicular
orientation. They strongly vary as function of electron energy. As
the angular distributions are created by multiple scattering and
interference, they depend on the electron wavelength. To separate
the effects caused by different electron wavelengths from effects
of the satellite excitation we have performed our experiment at
different photon energies, chosen such that the electron energy
for the line under investigation is always about 10~eV. We picked
the main line and the prominent lines 4, 5 and 7 to compare their
angular distributions in the molecular frame which are shown in
Figure 2. An electron energy of 10~eV is in the range of the
$\Sigma$ shape resonance for the main line
\cite{Shigemasa98prl,Jahnke02prl}. The angular distribution in
Fig. 2(a) for parallel orientation shows the well known
''guitar-like'' shape with the main lobe in the direction of the
Oxygen atom. Additional lobes at around 60 deg result from a
strong contribution of an $f$-wave which is characteristic for
this shape resonance. The two normal satellites 5 and 7 for which
the final state is of the same $\Sigma$ symmetry as the main line
show a qualitatively similar pattern. The angular distributions
show that the satellites exhibit a very similar shape resonance as
the main line, confirming similar claims of a shape resonance in
satellites made in
\cite{Schirmer91pra,reich94pra,ungier84prl,reimer86prl}. It is
worth noting that the molecular frame angular distribution of the
main line and the two satellites of the same symmetry are not
exactly equal. The multiple scattering is very sensitive to the
details of the potential and to vibrational excitation
\cite{Adachi93prl,Jahnke04bprl,Semenov04jpb,Semenov06jpb}. Both
are slightly different for the satellites, giving rise to the
variations in the angular distribution. Most interesting is the
angular distribution of peak number 4 (Fig. 2(d)) which is a
conjugate shake up to a state of $\Pi$ symmetry. In this case, as
well, the body fixed frame angular distribution is similar to the
main line, but in that case the angular distribution is plotted
for perpendicular orientation of the molecule. The shape of the
angular distribution found here strongly resembles the
characteristics of a $\Sigma$  shape resonance. The $\Sigma$,
however, refers to the symmetry of the continuum orbital. To reach
this orbital by conjugate shake up the polarization has to
perpendicular to the molecular axis. This shows that also
conjugate shake up satellites experience a $\Sigma$ shape
resonance at similar electron energies of the main line - for the
$\Pi$ orientation. This is in line with the generalization of the
atomic shake up model to states with molecular symmetry mentioned
above. The shake up matrix element is purely a (monopole) overlap
of the bound $4\sigma$ orbital to the $\epsilon\sigma$ continuum
state which forms the shape resonance. This can even be
rationalized approximately from atomic states which are by
multiple scattering converted to molecular continuum states with
higher angular momentum, since the bound $4\sigma$ orbital roughly
resembles an atomic $2p_z$ orbital. As the electron from this
orbital is shaken off it creates a $p_z$ wave in the continuum
very much like for the main line where this wave is created by
photoejection if the molecular axis is parallel to the
polarization.  This $p_z$ wave is subsequently multiply scattered
adding the higher angular momenta visible in \emph{guitar-like} shapes in
figure 2.

\section{Conclusions}
Our data give a very clear general trend: the molecular frame
angular distribution of the satellites in K-shell ionization of CO
are very similar to the angular distribution of the main line at
the same electron energy and at the same symmetry. I.e. a $\sigma$
like pattern as for the main line is found if molecule and
polarization are parallel. The same pattern evolves for satellites
where the remaining CO$^{+*}$(C1s$^{-1}$) has the same $\sigma$
symmetry as the CO$^+$(C1s$^{-1}$), while for satellites where the
CO$^{+*}$(C1s$^{-1}$) has $\Pi$ symmetry a strikingly similar
pattern arises if the polarization is perpendicular to the
molecular axis. As shape resonances manifest themselves nicely in
the body fixed frame angular distributions, the satellites show
equivalent shape resonances as the main line. These observations
can be reconciled with a two step model of photo absorption
followed by shake up and a multiple scattering of the continuum
electron in the molecular potential.

\acknowledgments This work was supported by DFG, the  Division of
Chemical Sciences,
 Geosciences and Biosciences Division, Office of Basic Energy Sciences, Office of Science,
 U. S. Department of Energy and the
Director, Office of Science, Office of Basic Energy Sciences and
Division of Materials Sciences under U.S. Department of Energy
Contract No. DE-AC03-76SF00098.  We are grateful for excellent
support during the beamtime by Elke Arenholz and Tony Young. We
are deeply indebted to an unknown referee of our paper, who has
pointed out a serious mistake in our interpretation and has
suggested the molecular version of the shake off as we
describe it now in the final version of this paper. \\

\end{document}